\documentclass{ws-ijmpc}
\usepackage[latin1]{inputenc}
\usepackage[english]{babel}
\usepackage{graphicx}
\usepackage{rotating}
\usepackage{wrapfig}
\usepackage[dvips]{epsfig}
\usepackage{amsmath}
\usepackage{amssymb}

\newlength{\epswidth}
\newcommand{\eq}[1]{Eq.\,(\ref{#1})}

\newcommand{\fig}[1]{Fig.\,\ref{#1}}

\newcommand{\bc}{\begin{center}}
\newcommand{\ec}{\end{center}}
\newcommand{\be}{\begin{equation}}
\newcommand{\ee}{\end{equation}}
\newcommand{\bea}{\begin{eqnarray}}
\newcommand{\eea}{\end{eqnarray}}
\newcommand{\ba}[1]{\begin{array}{#1}}
\newcommand{\Ref}[1]{\cite{#1}}
\newcommand{\Refs}[1]{Ref.\, \cite{#1}}
\newcommand{\ea}{\end{array}}
\newcommand{\angst}{\,\mathrm{\AA}}
\newcommand{\microns}{\mu\mathrm{m}}
\renewcommand{\epsilon}{\varepsilon}
\renewcommand{\theta}{\vartheta}
\renewcommand{\vec}[1]{\mathbf{#1}}

\newcommand{\pH}{p\mathrm{H}}
\newcommand{\pHvalue}{\mbox{$\pH$-value}}
\newcommand{\mmoll}{\mathrm{mmol/l}}
\newcommand{\invsec}{\mathrm{/s}}

\begin{document}

\markboth{M.~Hecht, J.~Harting, and H.J.~Herrmann}
{Formation and growth of clusters in colloidal suspensions}

\catchline{}{}{}{}{}

\title{Formation and growth of clusters in colloidal suspensions}

\author{Martin Hecht and Jens Harting}
\address{Institute for Computational Physics, Pfaffenwaldring 27, D-70569
Stuttgart, Germany\\
\{hecht,jens\}@icp.uni-stuttgart.de}

\author{Hans J. Herrmann}
\address{Computational Physics, IFB, Schafmattstr. 6, ETH Z\"{u}rich,
CH-8093 Z\"{u}rich, Switzerland\\ hans@icp.uni-stuttgart.de}

\maketitle

\begin{history}
\received{\today}
\revised{Day Month Year}
\end{history}

\begin{abstract}
Depending on the \pHvalue{} and salt concentration of Al$_2$O$_3$
suspensions different microstructures can form. Especially the clustered
one is of major interest for industrial purposes as found in the
production of ceramics. In this paper we investigate the clustered
microstructure by means of a coupled Stochastic Rotation Dynamics (SRD)
and Molecular Dynamics (MD) simulation. In order to gain statistics within
a system containing numerous clusters, large simulation volumes are
needed. We present our parallel implementation of the simulation algorithm
as well as a newly developed cluster detection and tracking algorithm. We
then show first results of measured growth rates and cluster size
distributions to validate the applicability of our method.

\end{abstract}

\ccode{PACS Nos.: 82.70.-y, 47.11.+j, 02.70.Ns}

\section{Introduction}
Colloidal suspensions in general are present everywhere in our daily life.
Paintings, cosmetic products, and different kinds of food are some
examples. They behave in a complex way, since different time and length
scales are involved. The particle sizes are on a mesoscopic length scale,
i.e., in the range of nanometers up to micrometers. Depending on the
particle sizes, materials, and concentrations, different interactions are
of relevance and often several of them are in a subtle interplay:
electrostatic repulsion, depletion forces, van der Waals attraction,
hydrodynamic interaction, Brownian motion, and gravity are the most
important influences. Here, we are interested in colloids, where
attractive van der Waals interaction is important for the description,
i.e., where under certain circumstances cluster formation plays an
important role\cite{Russel95,Hunter01}. To model these systems
experimentally, Al$_2$O$_3$ suspensions are commonly
used\cite{Richter03,Richter06}. Al$_2$O$_3$ is also a common material in
the ceramics industry. There, wet processing of suspensions, followed by a
sinter process is a common practice. The stability of the resulting
workpiece strongly depends on the properties of the clusters formed before
the sintering process. The size distribution, stability and local porosity
of the clusters as well as the time dependence of their formation are only
a few of the parameters of influence.

In our work we investigate cluster formation of a sheared solution
of spherical Al$_2$O$_3$ particles of diameter $0.37\,\mu$m in water. 
Cluster formation can have different reasons: depletion forces\cite{Ye96,Tuinier00,Rudhardt98}, like-charge attraction mediated
by the counterions in the solvent\cite{Tata92,Tata97,Linse99}, or,
as in our case, van der Waals attraction\cite{Huetter99,Wang99,Huetter00}.
The shear flow can either support cluster formation at low shear rates, 
or it can suppress cluster formation at high shear rates as we have shown 
in \Ref{Hecht06b}.
We adjust the simulation parameters so that the simulation corresponds
quantitatively to a real suspension with 5\% volume concentration under
shear. The shear rate is kept fixed at $\dot\gamma=20\invsec
$\cite{Hecht05,Hecht06}. For Al$_2$O$_3$ suspensions attractive van der
Waals forces compete with electrostatic repulsion. Depending on the
particle surface charge, clustering due to attractive van der Waals forces
can dominate or be prevented. We have presented how one can relate
parameters of DLVO potentials\cite{DLVO,DLVO2} with experimentally tunable
parameters, i.e., the \pHvalue{} and the salt concentration expressed by
the ionic strength $I$, influence the charge of the colloidal
particles\Ref{Hecht06}. We explored the stability diagram of Al$_2$O$_3$
suspensions and reproduced that the particles are uncharged close to the so
called ``isoelectric point'' at $\pH = 8.7$, where they form clusters
regardless of the ionic strength. For lower \pHvalue{}s particles can be
stabilized in solution. For very low \pHvalue{}s, low salt concentrations,
and high volume fractions a repulsive structure can be found. The particle
size is on a mesoscopic length scale, where Brownian motion is relevant
and long range hydrodynamic interactions are of importance. Therefore, we use
``Stochastic Rotation Dynamics'' (SRD), which includes both, hydrodynamics
and Brownian motion for the description of the fluid
solvent\cite{Malev99,Malev00}. 
\section{Simulation algorithm}
\noindent
Our simulation method is described in detail in\Ref{Hecht05,Hecht06} and
consists of two parts: a Molecular Dynamics (MD) code, which treats the
colloidal particles, and a Stochastic Rotation Dynamics (SRD) simulation
for the fluid solvent. In the MD part we include effective electrostatic
interactions and van der Waals attraction, known as DLVO
potentials\cite{DLVO,DLVO2}, a lubrication force and Hertzian contact
forces. DLVO potentials are composed of two terms, the first one being an
exponentially screened Coulomb potential due to the surface charge of the
suspended particles
\be
  V_{\mathrm{Coul}} =
  \pi \epsilon_r \epsilon_0
  \left[ \frac{2+\kappa d}{1+\kappa d}\cdot\frac{4 k_{\mathrm{B}} T}{z e}  
         \tanh\left( \frac{z e \zeta}{4 k_{\mathrm{B}} T} \right)
  \right]^2 \times \frac{d^2}{r} \exp( - \kappa [r - d]),
 \label{eq_VCoul}
\ee
where $d$ denotes the particle diameter, $r$ the distance between the
particle centers, $e$ the elementary charge, $T$ the temperature,
$k_{\mathrm{B}}$ the Boltzmann constant, and $z$ is the valency of the
ions of added salt. $\epsilon_0$ is the permittivity of the vacuum,
$\epsilon_r=81$ the relative dielectric constant of the solvent, $\kappa$
the inverse Debye length defined by $\kappa^2 = 8\pi\ell_BI$, with
ionic strength $I$ and Bjerrum length $\ell_B = 7\angst$. The first
fraction in \eq{eq_VCoul} is a correction to the DLVO potential (in the
form used in\Ref{Huetter99}), which takes the surface curvature into
account and is valid for spherical particles. The
effective surface potential $\zeta$ can be related to the \pHvalue{} of
the solvent with a $2pK$ charge regulation model\cite{Hecht06}. The
Coulomb term competes with the attractive van der Waals interaction
($A_{\mathrm{H}}=4.76\cdot 10^{-20}\,\mathrm{J}$ is the Hamaker constant)\cite{Huetter99}
\be
  V_{\mathrm{VdW}} = - \frac{A_{\mathrm{H}}}{12} 
     \left[ \frac{d^2}{r^2 - d^2} + \frac{d^2}{r^2} \right. 
             \;\left. + 2 \ln\left(\frac{r^2 - d^2}{r^2}\right) \right].
  \label{eq_VdW}
\ee
The attractive contribution $V_{\mathrm{VdW}}$ is responsible for the cluster
formation we observe. However, depending on the \pHvalue{} and the ionic strength,
it may be overcompensated by the electrostatic repulsion. When particles
get in contact, the potential has a minimum. However, \eq{eq_VdW} diverges 
due to the limitations of DLVO theory. We cut off the DLVO potentials and
model the minimum by a parabola. The particle contacts are modeled as Hertzian
contacts and for non-touching particles, below the resolution of the SRD 
algorithm short range hydrodynamics is corrected by a lubrication force, 
which we apply within the MD framework, as we have explained in \Refs{Hecht05,Hecht06}.
For the integration of translational motion of the colloidal particles we utilize a velocity Verlet
algorithm\cite{Allen87} and for the fluid we apply the Stochastic Rotation
Dynamics method (SRD)\cite{Malev99,Malev00}. It intrinsically contains
fluctuations, is easy to implement, and has been shown to be well suitable
for simulations of colloidal and polymer
suspensions\cite{Inoue02,Padding04,Gompper05,yeomans-2004-ali,Hecht05,Hecht06}.
The method is also known as ``Real-coded Lattice Gas''\cite{Inoue02} or as
``multi-particle-collision dynamics'' (MPCD)\cite{Gompper05} and is based
on coarse-grained fluid particles with continuous positions and velocities. A
streaming step and an interaction step are performed alternately. In the
streaming step, each particle $i$ is moved according to
$\vec{r}_i(t+\tau)=\vec{r}_i(t)+\tau\;\vec{v}_i(t)$,
where $\vec{r}_i(t)$ denotes the position of the particle $i$ at time $t$
and $\tau$ is the time step. In the interaction step fluid particles
are sorted into cubic cells of a regular lattice and only the particles
within the same cell interact according to an artifical collision rule
which conserved energy and momentum. First, for each independent cell $j$ the mean
velocity $\vec{u}_j(t')=\frac{1}{N_j(t')}\sum^{N_j(t')}_{i=1}
\vec{v}_i(t)$ is calculated. $N_j(t')$ is the number of fluid particles
contained in cell $j$ at time $t'=t+\tau$. Then, the velocities of each
fluid particle are rotated according to
\be
\label{eq_rotate}
\vec{v}_i(t+\tau) = \vec{u}_j(t')+\vec{\Omega}_j(t') \cdot [\vec{v}_i(t)-\vec{u}_j(t')].
\ee
$\vec{\Omega}_j(t')$ is a rotation matrix, which is independently chosen
at random for each time step and cell. Rotations are about one coordinate
axes by a fixed angle $\pm\alpha$.
To couple colloidal particles and the fluid, the particles
are sorted into SRD cells and their velocities are included in the
rotation step. The masses of colloidal and
fluid particles are used as a weight factor for the mean velocity 
\be
\label{eq_rotateMD}
\vec{u}_j(t')= \frac{1}{M_j(t')}\sum\limits^{N_j(t')}_{i=1}\vec{v}_i(t) m_i,
\qquad\mathrm{with}\qquad M_j(t')=\sum^{N_j(t')}_{i=1}m_i.
\ee
We sum over all colloidal and fluid particles in the cell to obtain their
total number $N_j(t')$. $m_i$ is the mass of particle $i$ and $M_j(t')$
gives the total mass contained in cell $j$ at time $t'=t+\tau$. We apply
shear by explicitly setting the mean velocity $\vec{u}_j$ to the shear
velocity in the cells close to the border of the system. A thermostat
removes the energy introduced to the system by the shear force.

A single simulation run as presented in our previous papers took between
one and seven days on a 3GHz Pentium CPU. However, for strongly clustering
systems we easily end up with only a single cluster inside the simulation
volume. In order to be able to gather statistics on cluster growth and
formation, as well as to minimize finite size effects, we parallelized our
code. While MD codes have been parallelized by many groups, only few
parallel implementations of a coupled MD and SRD program exist. This is in
contrast to the number of parallel implementations of other mesoscopic simulation
methods like for example the lattice Boltzmann method. A possible
explanation is that SRD is a more recent and so far not as widely used
algorithm causing the parallelization to be a more challenging task. In
order to push the development in this field we provide some details of our
implementation in this section.

We utilize the Message Pasing Interface (MPI) to create a C++ code based
on domain decomposition for both involved simulation methods. In the MD
code the position of neighbouring particles is needed to compute the
interactions. Since the intractions have a limited range, and a linked
cell algorithm is already used in the serial code, we apply linked cells
here as well. Particle positions at the border of the domain of each
processor are communicated to the neighbouring processors for the
calculation of the forces. Then, the propagation step is performed and
particle positions are updated, whereby the particles crossing a domain
boundary are transferred from one processor to the other one.

Since (in principle), fluid particles can travel arbitrary large distances
in one time step, one either has to limit the distance they can move, or
one needs all-to-all communication between the processors. Even though the
mean free path in our systems is small enough to limit communication to
nearest neighbours only, the current version of our code tries to be as
general as possible and allows fluid particles to move to any possible
position in the total simulation volume within a single timestep. First,
we determine locally which fluid
particles have to be sent to which destination CPU and collect all particles to
be sent to the same destination into a single MPI message. If no particles are
to be sent, a zero dummy message is transmitted. On the receiving
side, \verb+MPI_Probe+ with the \verb+MPI_ANY_SOURCE+ option is utilized to
determine the sender's rank and the number of particles to be accomodated.
Now, \verb+MPI_Recv+ can be used to actually receive the message. All
processors send and receive in arbitrary order, thus waiting times are kept at
a minimum allowing a very efficient communication. The standard MPI all-to-all
communication procedure should be less efficient since the size of every
message would be given by the size of the largest message. However, we still
do find a substantial communication overhead from our benchmark tests of the
scalability of the code. Due to this overhead, we are currently limited to
32-64 CPUs on an IBM p690 cluster. In order to achieve Gallilean invariance, a
random shift of the SRD lattice is performed for every rotation
step\cite{Ihle02a,Ihle02b}. Since the domains managed by each CPU do not move,
this would include the borders between the processors to cross SRD cells, which
is undesirable. Therefore, we keep the position of the lattice fixed and shift
the fluid particle positions before sorting them into the cells instead. After
the rotation step they are shifted back.

\section{Results}
\begin{figure}
\qquad a)\qquad $t=0.05$s\hspace{3.5cm}b)\qquad $t=0.26$s
\vspace{-0.3cm}
\bc
\epsfig{file=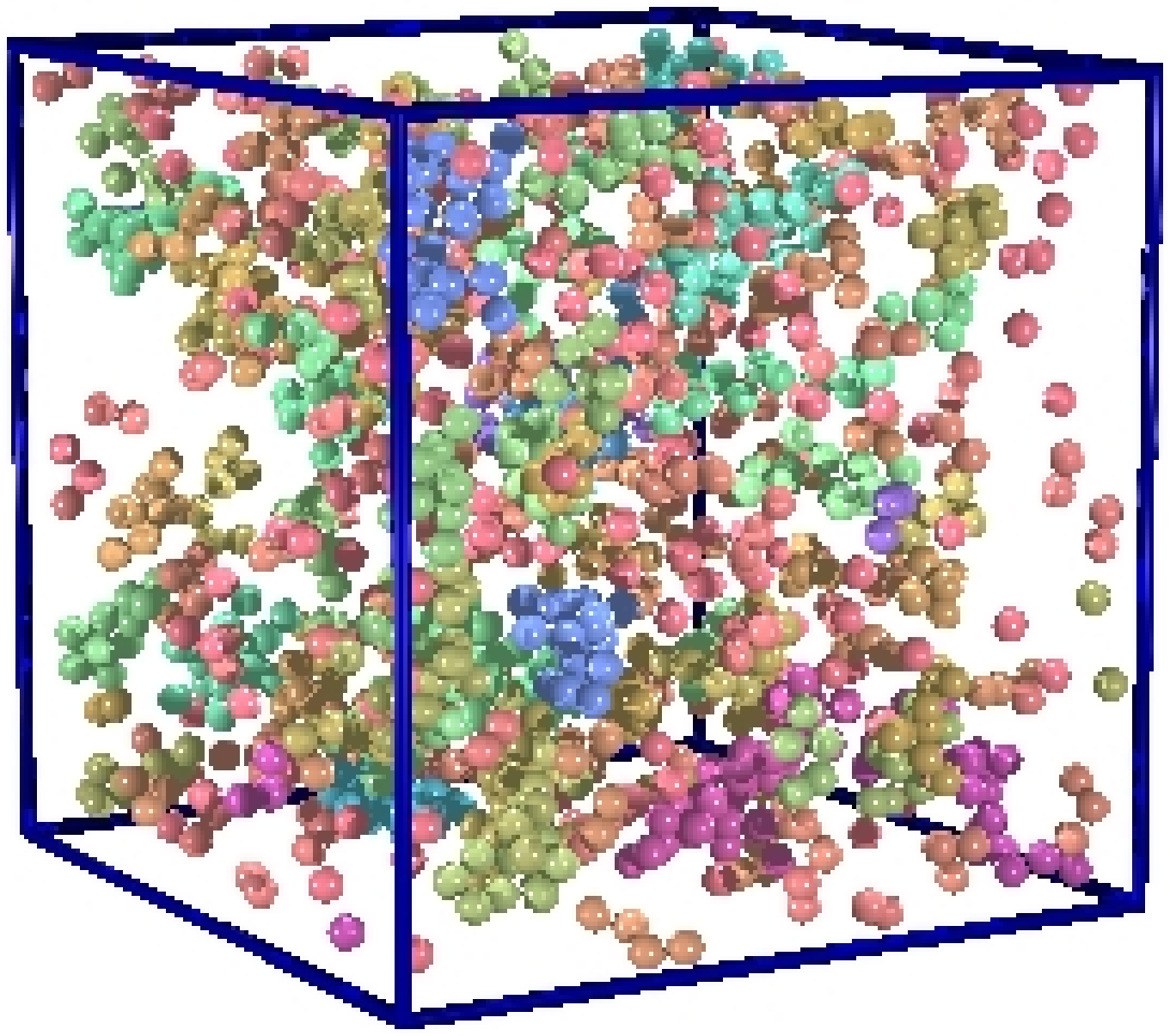,width=0.6\epswidth}
\epsfig{file=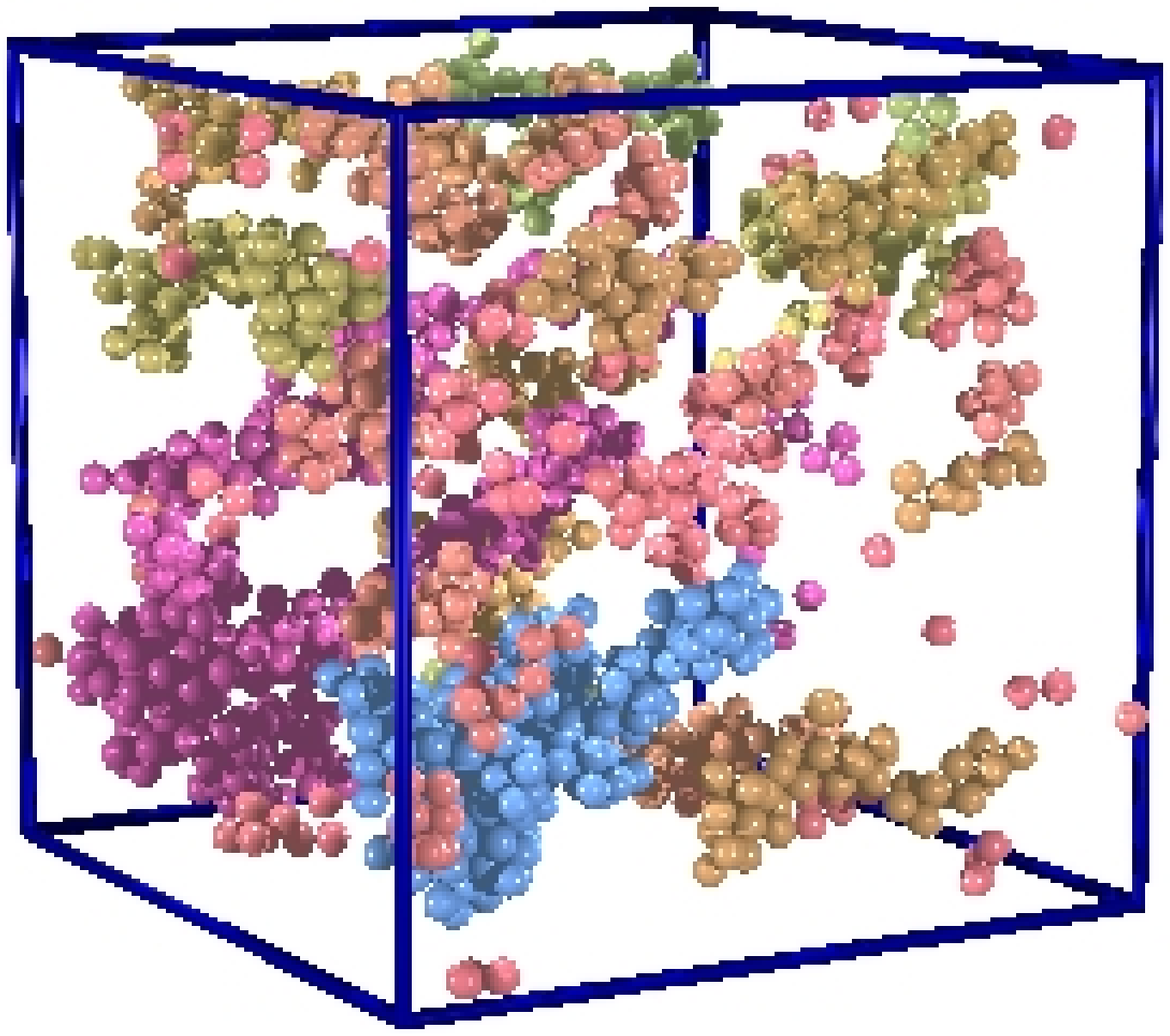,width=0.6\epswidth}
\ec
\qquad c)\qquad $t=1.06$s\hspace{3.5cm}d)\qquad $t=4.22$s
\vspace{-0.3cm}
\bc
\epsfig{file=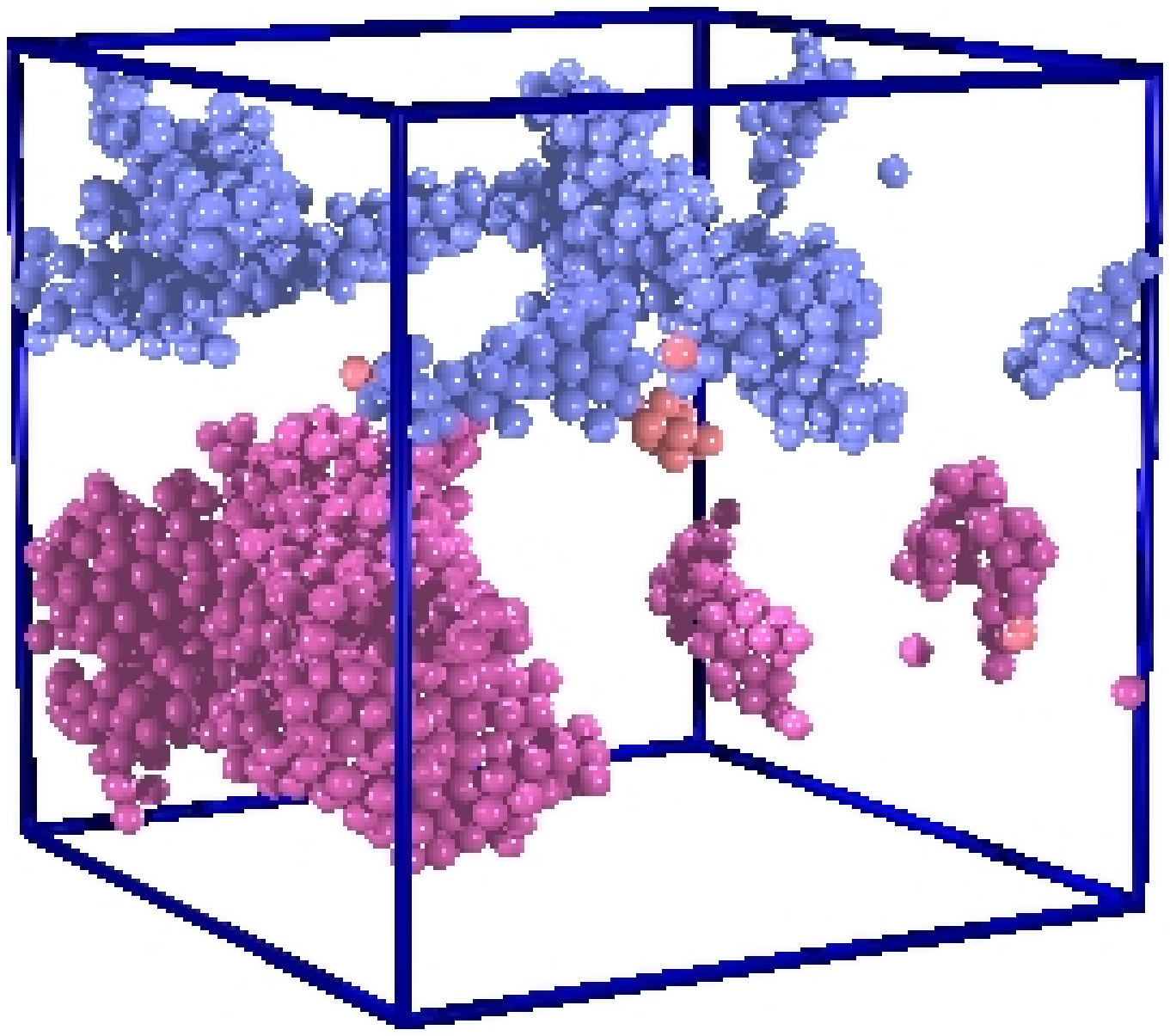,width=0.6\epswidth}
\epsfig{file=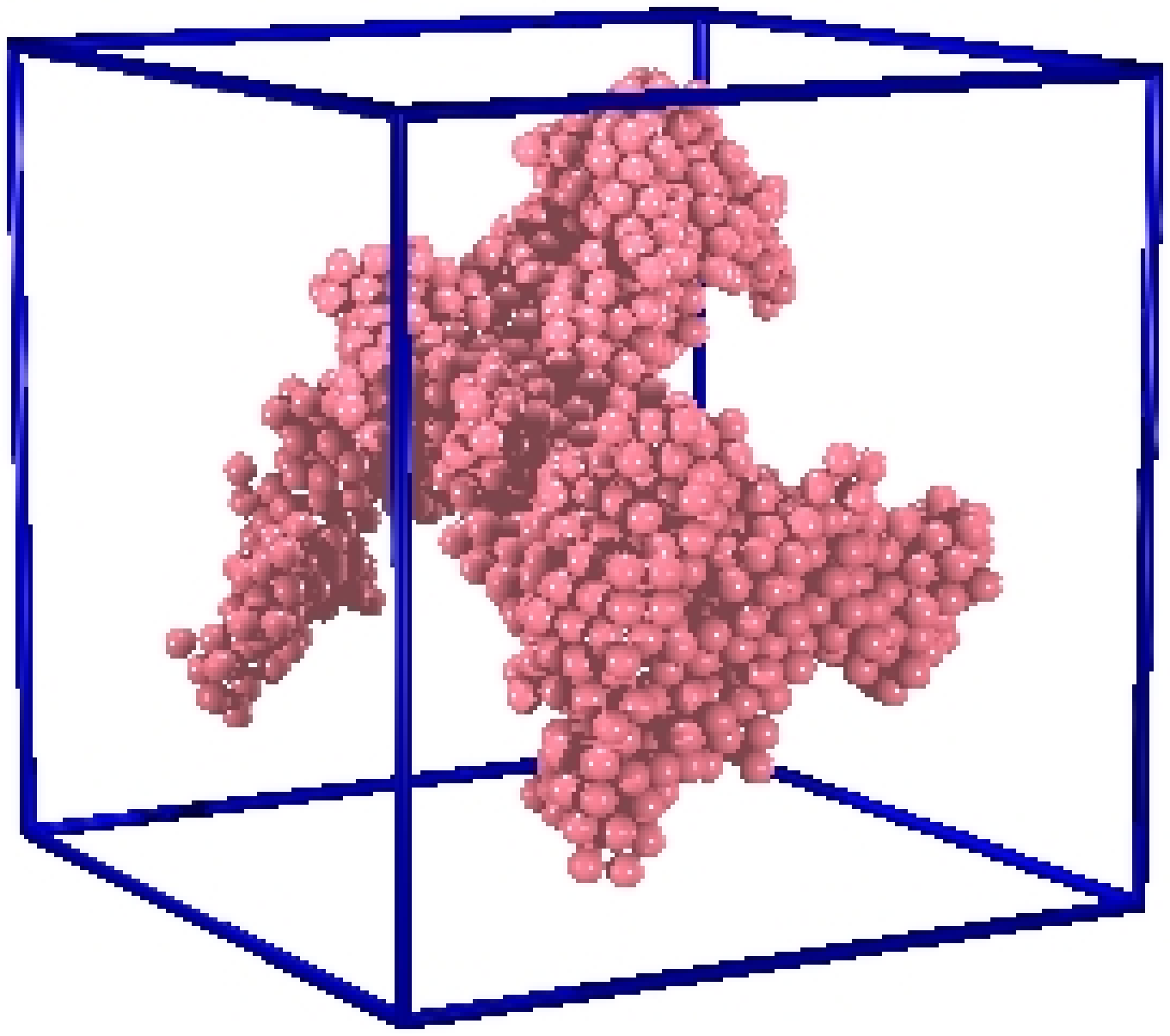,width=0.6\epswidth}
\ec
\caption{Snapshots of a simulation of an $8.88\microns^3$ system,
filled with $\Phi=5\%$ MD particles of
diameter $d=0.37\microns$ under shear with $\dot\gamma=20\invsec$.}
\label{fig_snapshots}
\end{figure}

We study the formation of clusters for systems containing a volume
concentration of $5\%$ of colloidal particles (=1320 MD particles), a shear rate of
$\dot\gamma=20\invsec$, ionic strengths $I = 3\mmoll$ and $7\mmoll$, and
$\pH = 6$ and $7$. To demonstrate the effect of clustering, in
\fig{fig_snapshots} snapshots from a typical simulation of a
$8.88\microns^3$ system with periodic boundaries at $I = 7\mmoll$ and $\pH = 6$ at different times
are shown.  While at the beginning of the simulation (a), freely moving
particles can be observed, small clusters appear after $t=0.26$s (b).
After $t=1.06$s, all particles are contained within three individual
clusters (c) and after $t=4.22$s only a single cluster is left in the
system. For an investigation of the formation and movement of clusters,
substantially larger systems are needed. Therefore, we scale up the
simulation volume to $17.76\microns^3$ containing 10560 MD particles and
1.3$\cdot$10$^7$ fluid particles. Due
to the computational demands of the fluid solver, a single simulation of
$5$s real time requires about 5000 CPU hours on 32 CPUs of an IBM p690
system.

We developed a cluster detection algorithm which not only examines a
certain configuration at a fixed time, but also takes account of the time
evolution of clusters. This algorithm works as follows: a cutoff radius is
introduced, below which two particles are considered to be connected. If
they are separated further, they are considered as being not directly
connected. However, they might both be connected to a third particle.
Therefore we have to check all particle pairs for possible connections. If
there are no further connections, the particles are considered not to
belong to any cluster. Otherwise four cases have to be distinguished:
If both particles are not part of any cluster, a new cluster is created
and both particles are assigned to it (1). If one particle is already part
of a cluster, the other one is assigned to the same cluster (2). If both
particles belong to different clusters, the clusters are united, i.e., all
particles of the smaller cluster are assigned to the larger one (3). If it
is found that for a particle pair to be checked later on, both particles
already belong to the same cluster, nothing has to be done (4).
This pairwise checking is optimized by a linked cell algorithm, so that
only particle pairs of the same and of neighbouring cells are checked.
Additionally, clusters need to be tracked in time, i.e., the clusters found
within a time step have to be identified with the clusters of the previous
time step. This is done by assigning an identification number (``cluster
ID'') for every cluster. Since every particle has a unique identification
number, assigning the ID of the cluster it belongs to solves the problem.
According to which cluster ID the particles were assigned in the previous
time step, the ID is assigned to the new cluster. Again four different
cases have to be considered:
if both particles belonged to the same cluster, and therefore refer to the same
cluster ID, this ID is assigned to the new cluster (1).
If one of the particles did not belong to any cluster in the previous time
step, a new cluster has formed during the last time step and has to be provided
with a new ID (2).
If only one particle was part of a cluster, the ID it provides is preserved (3).
If the particles are assigned to different cluster
IDs one of those IDs has to be choosen for the new cluster. We decide for
the one referring to the larger cluster of the previous time step or choose randomly if both clusters are of identical size (4).
Finally one has to check if the cluster IDs are unique. If several clusters are
assigned to the same id the largest one keeps the ID and the smaller ones are
assigned to new ones.

\begin{figure}
\bc
\epsfig{file=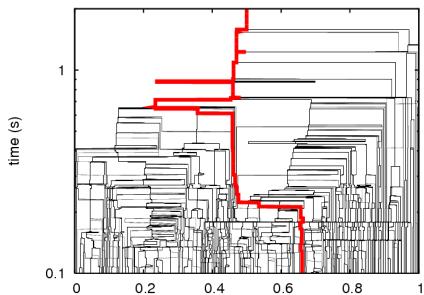,width=0.62\epswidth}
\ec
\caption{Tree-like structure of the assignment of colloidal particles to
individual clusters. This cluster tree corresponds to the simulation
presented in \fig{fig_snapshots}. For better visibility, the time is given
on a logarithmic scale. All particles start at the bottom of
the plot and aggregate in individual clusters as shown by the merging
lines. After two seconds, only a single large cluster is left
in the system. The thick line denotes the path of a single particle in order to
demonstrate the possibility to track which cluster this particle
belongs to and how these clusters break up and remerge.
}
\label{fig_clustertree}
\end{figure}

The strength of our algorithm is the possibility to track individual particles
and their assignment to different clusters in time. Clusters grow and break
into pieces and we can follow the trajectory of each particle in this scenario.
This enables us to draw cluster assignment trees like the one in
\fig{fig_clustertree}. In contrast to conventional algorithms, where 
clusters cannot be tracked in time, the clusters are sorted 
here on the $x$ axis and keep their position. The lines are 
obtained by plotting the assignment of the particles to the clusters and
their distance depicts the cluster sizes, i.e., if at time
$t$ a cluster contains a
fraction $p$ of all particles in the simulation, a fraction $p$ is reserved for this cluster on the  $x$-axis and the line is plotted
at the center of this region. Consequently, if only a single cluster is
left in the system, the corresponding line is drawn at $x=0.5$.
Depending on the inter particle forces, different structures can be
identified, meaning different scenarios like breaking up of large clusters
or unification of smaller ones. We are planning to study systematically
the dependence of the structures seen in such cluster tree plots on the
inter particle forces determined by the \pHvalue{} and the ionic strength
$I$ in a future work.

\begin{figure}
a)\hspace{6cm} b)
\vspace{-0.4cm}
\bc
\epsfig{file=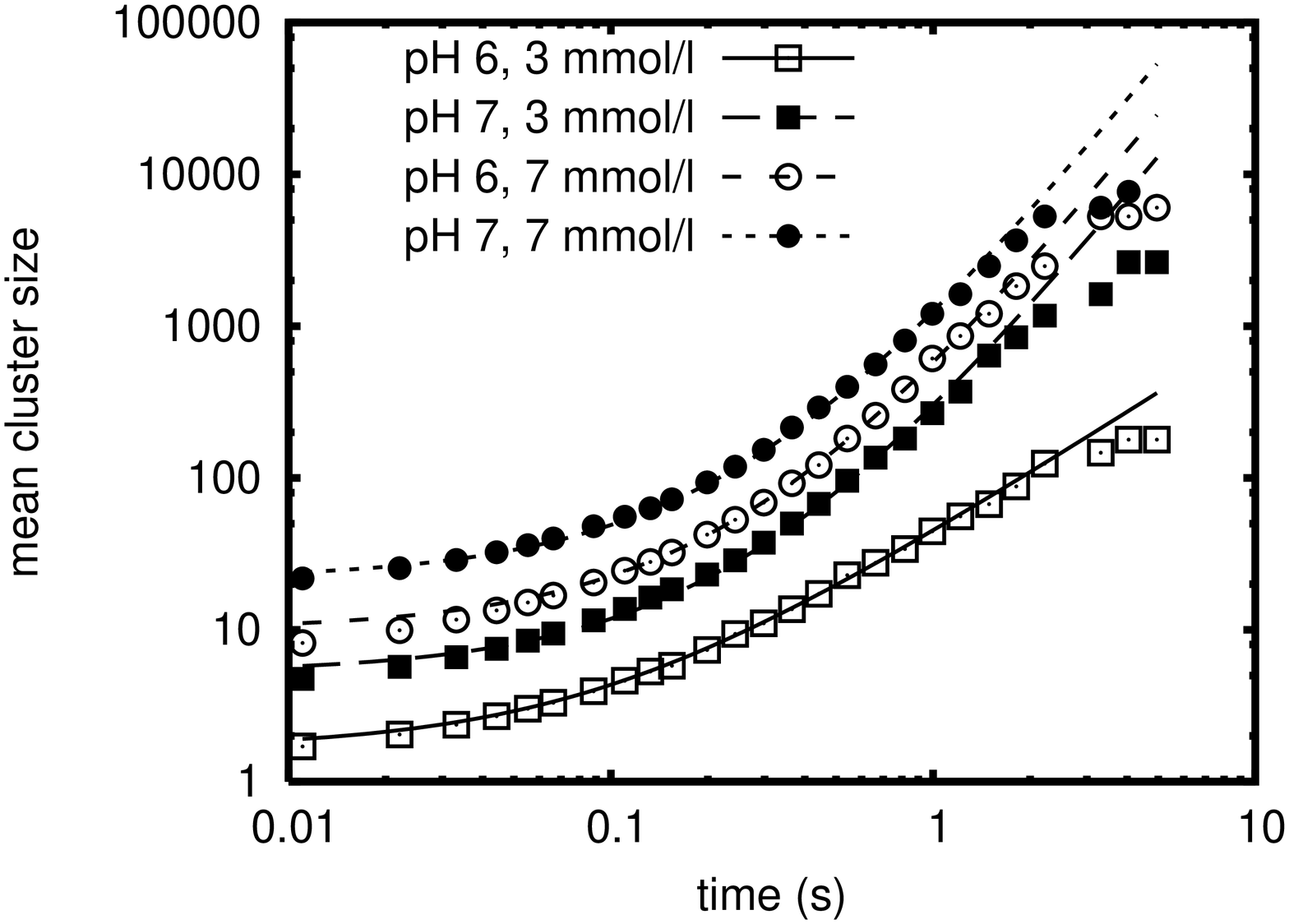,width=0.65\epswidth}
\epsfig{file=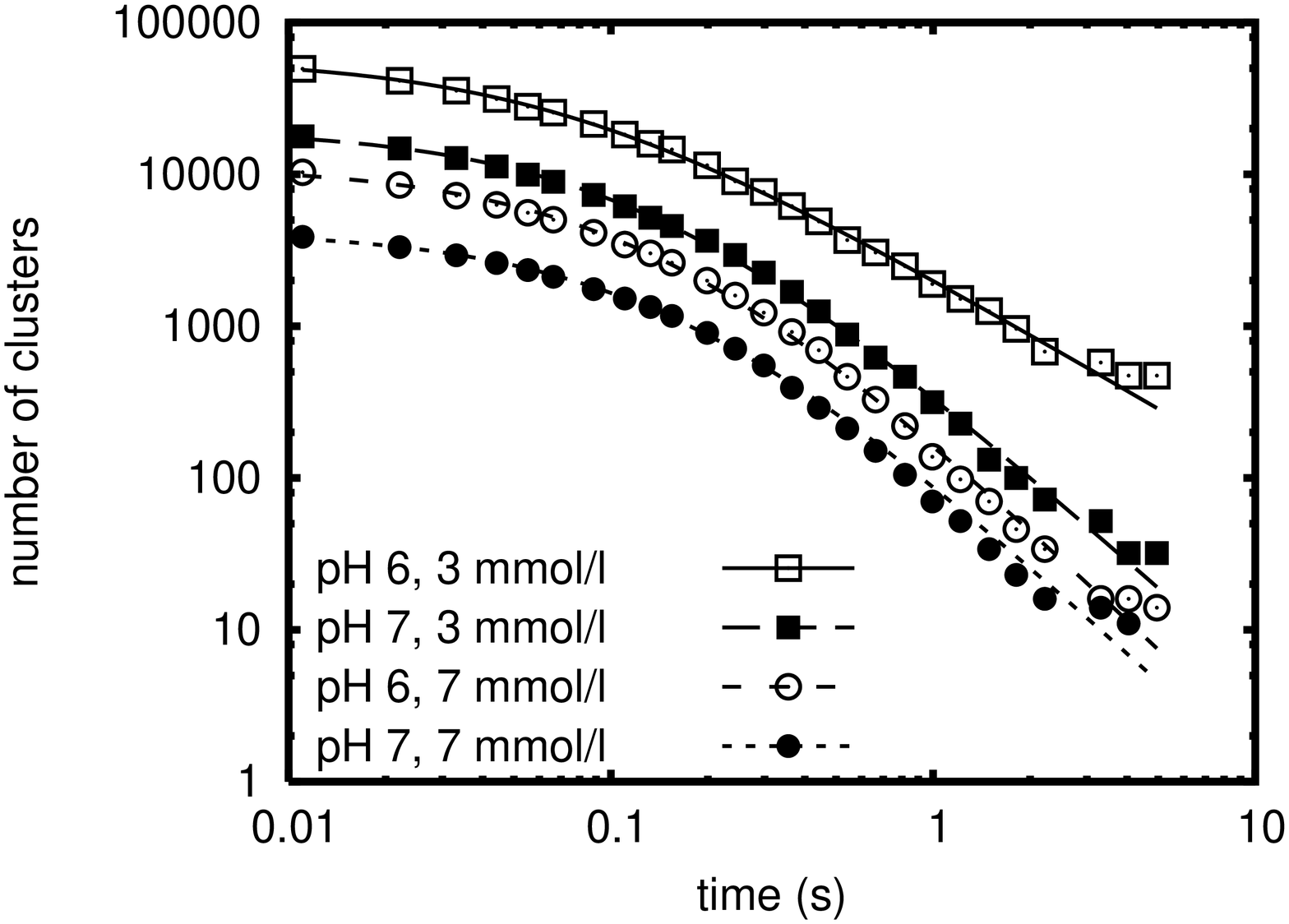,width=0.65\epswidth}
\ec
\caption{The time dependence of the mean cluster size 
is plotted for different simulation parameters (a).
Fig.~b) depicts the time dependence of the number of clusters found
in the system. Each curve is shifted vertically
by a factor of two for better visibility. While the symbols
correspond to simulation data, lines are given by a power law fit.}
\label{fig_clustersizes}
\end{figure}

In \fig{fig_clustersizes} we present the time dependence of the mean
cluster size (a) and of the number of clusters in the system (b). We find that
both observables can be fitted by a power law of the form
$A\cdot(t+B)^C$, where $A,B,C$ are fitting parameters. The lines in the
figure correspond to the fit and the symbols to the simulation data. 
The parameters $A,B,C$ to fit the simulation data shown in \fig{fig_clustersizes}
are listed in table\,\ref{tab_fitparams}. 
It would be of great interest to investigate if a general scaling behavior can be
observed depending on the volume concentration, the ionic strength and the
pH value. However, for this a detailed investigation of the parameter
space would be needed which will be the focus of a future work.

\begin{table}
\bc
\begin{tabular}{|c|c|c|c|c|c|c|c|}
\hline
\multicolumn{2}{|c|}{conditions} &
\multicolumn{3}{c|}{number of clusters} &
\multicolumn{3}{c|}{mean cluster size} \\ \hline
$\pH $ & $\frac{I}{\mathrm{mmol/l}}$ & $A$ & $B/\mathrm{ms}$ & $C$ %
& $A$ & $B/\mathrm{ms}$ & $C$ \\
\hline
6 & 3 & $1.5\cdot 10^6$   & 71 & -1.25 & $3.27\cdot 10^{-3}$ &  96 & 1.36\\ \hline
7 & 3 & $5.27\cdot 10^7$ & 131 & -1.9  & $8.42\cdot 10^{-7}$ & 277 & 2.66\\ \hline
6 & 7 & $1.46\cdot 10^8$ & 142 & -2.047 & $8.05\cdot 10^{-7}$ & 277 & 2.66\\ \hline
7 & 7 & $1.01\cdot 10^8$ & 162 & -1.98 & $8.72\cdot 10^{-7}$ & 277 & 2.66\\ \hline
\end{tabular}
\ec
{\small Table 1. Parameters for the fit of the simulation data}
\label{tab_fitparams}
\end{table}


\section{Conclusion}
In this paper we have demonstrated an efficient way to
parallelize a combined SRD and MD code and presented our new cluster
detection algorithm that is able to not only detect clusters, but also to
track their positions in time. We applied this algorithm to data
obtained from large scale simulations of colloidal suspensions in the
clustering regime and find that the time dependence of the mean cluster
size and the number of clusters in the system can be well described by
power laws.

\section*{Acknowledgments}
This work has been financed by the German Research Foundation (DFG) within the
project DFG-FOR 371 ``Peloide''. We thank G.~Gudehus, G.~Huber, M.~K\"ulzer,
L.~Harnau, M.~Bier, J.~Reinshagen, S.~Richter, and A.~Coniglio for valuable
collaboration. We also thank the German-Israeli Foundation (GIF) for support. The
computations were performed on the IBM p690 cluster at the Forschungszentrum
J\"{u}lich, Germany and at HLRS, Stuttgart, Germany. 

\appendix


\end{document}